\documentclass[pra, showpacs, floatfix]{revtex4}
\usepackage{graphicx}
\usepackage{amsmath, amsfonts, amssymb, bm}

\begin{document}
\title{Strong-field spatial intensity-intensity correlations of light scattered from regular structures of atoms}

\author{Mihai~\surname{Macovei}}
\email{mihai.macovei@mpi-hd.mpg.de}

\author{J\"org~\surname{Evers}} 
\email{joerg.evers@mpi-hd.mpg.de}

\author{Christoph~H.~\surname{Keitel}}
\email{keitel@mpi-hd.mpg.de}

\affiliation{Max-Planck-Institut f\"ur Kernphysik, 
 Saupfercheckweg 1, D-69117 Heidelberg, Germany}

\date{\today}
\begin{abstract}
Photon correlations and cross-correlations of light 
scattered by a regular structure of strongly driven
atoms are investigated. At strong driving, the scattered
light separates into distinct spectral bands, such that 
each band can be treated as independent, thus extending
the set of observables. We focus on second-order 
intensity-intensity correlation functions in two- and multi-atom
systems. We demonstrate that for a single two-photon detector 
as, e.g., in lithography, increasing the driving 
field intensity leads to an increased spatial resolution 
of the second-order two-atom interference pattern. 
We show that the cross-correlations between photons emitted
in the spectral sidebands violate Cauchy-Schwartz 
inequalities, and that their emission ordering cannot be
predicted. Finally, the results are generalized for
multi-particle structures, where we find results different
from those in a Dicke-type sample.
%
\end{abstract}

\maketitle
\section{Introduction}
Throughout the history of quantum physics,
first-order correlations have played a major role
in discussing the foundations of the underlying 
theory. One of the most famous model system is
Young's double-slit, which despite its simplicity
allows to explore fundamental questions such as
complementarity and uncertainty 
relations~\cite{ficek,uncert,uncert2,uncert3}.
A modern realization of Young's experiment involves
two atoms scattering near-resonant laser
light~\cite{eichmann,mirror,int1,int,int2,int3}.
The light is thus scattered by the simplest form
of a regular structure, which gives rise to interference 
phenomena in the scattered light, because different
indistinguishable pathways connect source and 
detector~\cite{BornWolf}. Analogous interference
is also possible with single particles, demonstrating
pathway interference in the energy-time 
domain~\cite{otherschemes}.

Starting with the experiments by Hanbury-Brown and
Twiss~\cite{hbt} involving the measurement on a
thermal field and the corresponding measurement with
anti-bunched resonance fluorescence light~\cite{antibunch},
the second-order correlation function gained considerable
attention. This correlation function measures
intensity correlations and thus provides information
on the fluctuation of the fields~\cite{g2}.

In general, spatial interference in the different correlation
functions can only be expected under certain experimental
conditions. For example, it is possible to have
interference in the second-order correlation function
under conditions of no interference of the first-order
correlation function~\cite{int3}.
In particular the first order interference in light
scattered of regular structures of atoms is typically
restricted to low incident light intensity and vanishes 
at strong driving~\cite{int1,int,int2,int3,cbs,vogel,letter}. 
For example, in a two-particle system in the strong 
field limit, the collective dressed states 
are uniformly populated, and thus the 
interference fringe visibility is zero. 
Since applications often rely on these interference
effects~\cite{cbs,vogel,double-res,letter}, this restricts 
the potential implementation, where properties such as 
coherence, increased resolution,
high signal-to-noise ratios or a rapid coherent system 
evolution are among the desirable properties of
the scattered light.

More fundamentally, different lines of the frequency
spectrum of the scattered light become separated
in the strong-field limit, just as
in the Mollow resonance fluorescence spectrum of a single 
strongly-driven two-level atom. We have shown
recently that this separation can be employed
to recover full first-order interference fringe
visibility in the strong-field limit, and to
gain a clearer interpretation and an extended 
set of observables to describe the scattered 
light~\cite{letter}. The basic idea was
to facilitate a frequency-dependent electromagnetic
bath, which can be realized, for example, using 
cavities or photonic crystals~\cite{Cexp,John}. The focus of this 
work, however,
was on first-order correlation functions.

Here, we discuss spatial second-order quantum 
interference effects in strong driving fields. 
As in the strong-field limit the spectral lines are 
well-separated, we define observables for each of 
the spectral lines separately.
This allows for additional observables such as
cross-correlations between different the spectral lines.
First, we investigate the spatial dependence of the 
second-order correlation functions for a strongly driven 
atomic pair in free space and focus on the 
case of a single two-photon detector registering photons 
at the driving laser frequency $\omega_{L}$, as, for 
example, in lithography with a medium sensitive to two-photon 
exposure. We show that in this setup, the spatial resolution 
of the central strong-field second-order interference pattern 
can be increased by a factor of two as compared to the 
corresponding weak-field pattern simply by increasing the 
driving field intensity.
This allows to create structures with high 
spatial resolution and signal intensity.
Next, quantum cross-correlations between photons emitted 
in different spectral bands are investigated. In particular, 
we show that the spatial Cauchy-Schwarz inequalities (CSI) 
are violated for photons emitted into the sideband spectral 
lines over a wide range of detector positions. 
Also, it is impossible to predict 
the temporal ordering of two photons emitted 
in different sidebands.
Our scheme can be realized in a wide range of 
systems, and can also be employed to analyze the structure 
of the scatterers. Finally, we generalize 
our results to the case of a linear chain of $N$ atoms,
where the results are found to be different from
results  in Dicke-type samples.

\begin{figure}[b]
\begin{center}
\includegraphics[width=14cm]{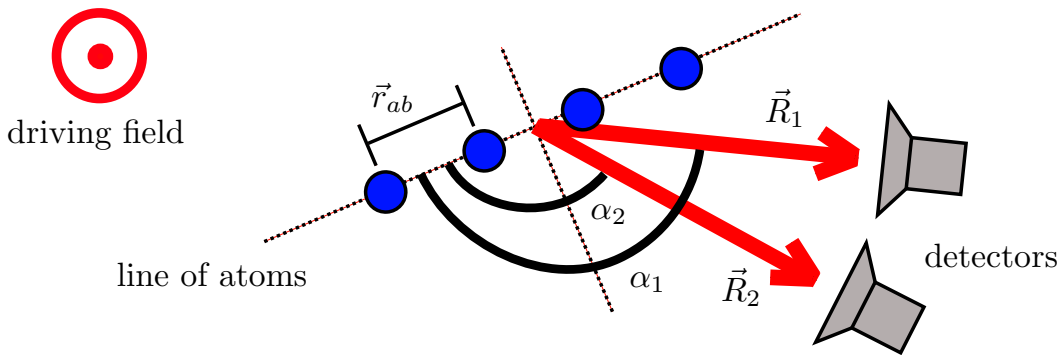}
\end{center}
\caption{\label{fig-1} A linear structure of two-state emitters,
where neighboring atoms are separated by the distance vectors 
$\vec r_{ab}$. The whole structure is driven 
by a resonant strong external laser field with wave vector 
$\vec k_{L}$. The two detectors are used to measure correlations 
among the photons emitted in observation directions  $\vec R_{1}, \vec R_{2}$,
which are also characterized by the angles $\alpha_{1}, \alpha_{2}$ between 
$\vec r_{ab}$ and $\vec R_{1}, \vec R_{2}$.
The line of atoms and the two detectors 
are located in a plane, and the driving field propagates perpendicular
to this plane.}
\end{figure}

\section{Analytical treatment}
We first concentrate on a driven pair of distinguishable non-overlapping 
two-state emitters, labeled $a$ and $b$, and  
generalize the discussion to a linear structure of $N>2$ atoms in 
section~\ref{multi}.
The two atoms both have atomic transition 
frequency $\omega_{0}$, are located at positions $\vec{r}_a$, $\vec{r}_b$,
and are separated by $\vec r_{ab}$. The electronic bare 
ground state of atom $j\in\{a,b\}$ is labeled
by $|1\rangle_j$, the corresponding excited state is $|2\rangle_j$.
The external driving field with frequency 
$\omega_{L}= ck_{L}$ and wave vector $\vec k_{L}$  is aligned such that 
$\vec k_{L}\cdot \vec r_{ab}=0$ (see Fig.~\ref{fig-1}). The particles 
are spontaneously damped via the interaction with the surrounding 
electromagnetic field (EMF) reservoir. Our aim is to 
investigate the spatial photon correlations of the light scattered by 
the atoms in the intense driving field limit. 

Assuming the electric dipole and
rotating wave approximations, the laser-dressed atomic system is described
by the Hamiltonian $H$ = $H_{0}$ + $H_{I}$, with
\begin{align}
H_{0} &= \sum_{k}\hbar(\omega_{k}-\omega_{L})a^{\dagger}_{k}a_{k} 
+ \sum_{j\in \{a,b\}}\hbar \tilde \Omega R_{zj}, \label{H0}  
\allowdisplaybreaks[2]\\
H_{I} &= i\sum_{k}\sum_{j \in \{a,b\}}(\vec g_{k}\cdot \vec d_{j})
\left \{ 
a^{\dagger}_{k} \left (R_{zj}\frac{\sin{2\theta}}{2}  
- R^{(j)}_{21}\sin^{2}{\theta}
+ R^{(j)}_{12}\cos^{2}{\theta} \right )
e^{-i(\vec k - \vec k_{L})\cdot \vec r_{j}} - \textrm{H.c.} \right \}. 
\label{HI} 
\end{align}
Here, $H_{0}$ is the Hamiltonian of the free EMF and the free dressed 
atomic subsystems, while $H_{I}$ describes the
interaction of the laser-dressed atoms with the EMF.  
$a_{k}$ and $a^{\dagger}_{k}$ are the radiation field annihilation 
and creation operators obeying the commutation relations
$[a_{k},a^{\dagger}_{k^{'}}]=\delta_{kk^{'}}$, 
and $[a_{k},a_{k^{'}}]$= $[a^{\dagger}_{k},a^{\dagger}_{k^{'}}]=0$. 
The atomic operators $R^{(j)}_{\alpha \beta}=|\tilde 
\alpha\rangle_{j} {}_{j}\langle \tilde \beta|$ describe the transitions 
between the dressed states $|\tilde \beta \rangle_{j}$ and $|\tilde 
\alpha \rangle_{j}$ in atom $j$ for $\alpha \not= \beta$ and 
dressed-state populations for $\alpha=\beta$, and satisfy the 
commutation relation 
\begin{equation}
[R^{(j)}_{\alpha\beta},
R^{(l)}_{\alpha^{'}\beta^{'}}]=\delta_{jl}
\left [\delta_{\beta \alpha^{'}}R^{(j)}_{\alpha \beta^{'}}-\delta_{\beta^{'} 
\alpha}R^{(j)}_{\alpha^{'}\beta} \right]\,.
\end{equation}
The operators $R^{(j)}_{\alpha \beta}$ 
can be represented through the bare state operators via the 
transformations 
\begin{align}
|1\rangle_{j} = \sin{\theta}|\tilde 2\rangle_{j}+\cos{\theta}|\tilde 1\rangle_{j} \,,
\qquad
|2\rangle_{j} =\cos{\theta}|\tilde 2\rangle_{j}-\sin{\theta}
|\tilde 1\rangle_{j}\,,
\end{align}
where the mixing angle $\theta$ is given by $\cot{2\theta} = \Delta/(2\Omega)$. 
The laser field
detuning is $\Delta = \omega_{0} - \omega_{L}$, and the Rabi frequency
is defined by $2\Omega=(\vec d\cdot \vec E_{L})/\hbar$. Here, $E_{L}$ is the electric 
laser field strength, and $\vec d \equiv \vec d_{a} = \vec d_{b}$ is the 
transition dipole matrix element. 
We further define
the population inversion operators $R_{zj}=|\tilde 2\rangle_j {}_j\langle 
\tilde 2| - |\tilde 1\rangle_j {}_j\langle \tilde 1|$ and the generalized
Rabi frequency 
$\tilde \Omega=
[\Omega^{2}+(\Delta/2)^{2}]^{1/2}$.
The dressed state transition frequencies are $\omega_L, 
\omega_\pm = \omega_L \pm 2\tilde{\Omega}$. 
The two-particle 
spontaneous decay and the vacuum-mediated collective interactions are 
given by the frequency-dependent expression 
\begin{eqnarray}
\gamma_{jl}(\omega) = \gamma(\omega)[\chi_{jl}(\omega)+i\Omega_{jl}(\omega)]\,.
\end{eqnarray}
Independent of the atom-vacuum coupling, the collective parameters $\chi_{jl}$ and 
$\Omega_{jl}$ ($j\neq l$) tend to zero in the large-distance case 
$r_{jl} \to \infty$ which  corresponds to the absence of coupling 
among the emitters. In the small-distance case $r_{jl} \to 0$, the parameter $\chi_{jl}$ 
tends to unity, while $\gamma\Omega_{jl}$ tends to 
the static dipole-dipole interaction potential. 

In the intense field limit $\tilde \Omega \gg N\gamma$, different lines of the spectrum 
are well-separated. This allows us to investigate the spatial dependence of the 
second - order correlation functions for each of the spectral band separately. In this 
case, it follows from the interaction Hamiltonian Eq.~(\ref{HI}) that the operators 
\begin{eqnarray}
R_{zj}\sin{2\theta}/2, \qquad R^{(j)}_{21}\cos^{2}\theta \qquad {\rm and}
~~R^{(j)}_{12}\sin^{2}\theta 
\label{s_op} 
\end{eqnarray}
can be considered as the sources of the $m$th spectral field components 
$\{m \in C,R,L\}$. Here, $C$ stands for the central spectral band 
emitted at frequency $\omega_{L}$, and $R$ and $L$ are for the right and 
left sidebands emitted at frequencies $\omega_{L}+2\tilde \Omega$ and 
$\omega_{L}-2\tilde \Omega$, respectively~\cite{KL}. 
In the following, we use this decomposition to investigate the spatial properties
of the photon statistics 
and cross-correlations between photons emitted from different spectral lines.

\section{Two-particle quantum dynamics}
In order to be able to derive analytic expressions for the required expectation
values, we first investigate the two-atom quantum dynamics 
in a strongly driven laser field. We introduce, for this reason, the two-atom 
collective dressed states as 
\begin{align}
|\Psi_{e}\rangle = |\tilde 2_{a},\tilde 2_{b}\rangle 
\,, \quad
|\Psi_{s(a)}\rangle = \frac{1}{\sqrt{2}} (|\tilde 2_{a},\tilde 1_{b}\rangle \pm |\tilde 2_{b},
\tilde 1_{a}\rangle )
\,, \quad \textrm{ and } |\Psi_{g}\rangle=|\tilde 1_{a},
\tilde 1_{b}\rangle \,.
\end{align}
The variables $\sigma_{\alpha \beta}=\langle |\Psi_{\alpha}
\rangle \langle \Psi_{\beta}|\rangle$ are expectation values of the corresponding transition
($\alpha \not = \beta$) and population operators ($\alpha = \beta$)
$(\{\alpha,\beta \} \in \{e,s,a,g \})$. Using Eqs.~(\ref{H0},\ref{HI}), one can
obtain the equations of motion for the dressed-state variable of interest, i.e.,
$x=2(\sigma_{ee}-\sigma_{gg})$, $y=\sigma_{ss}-\sigma_{aa}$, and 
$z=\sigma_{ee}+\sigma_{gg}-\sigma_{ss}-\sigma_{aa}$, as~\cite{mario}:
\begin{subequations}
\label{twoEq}
\begin{align}
\dot x(t) &= -2\xi^{(+)}x + 4\zeta^{(-)}_{ab}y + 4 \xi^{(-)} \,, 
 \allowdisplaybreaks[2]\\
\dot y(t) &= -\zeta^{(-)}_{ab}x - 2 (c^{(0)}_{ab} + \xi^{(+)})
y + 2\zeta^{(+)}_{ab}z \,, 
 \allowdisplaybreaks[2] \\
\dot z(t) &= 2\xi^{(-)}x + 4\zeta^{(+)}_{ab}y - 4\xi^{(+)}z \,. 
\end{align}
\end{subequations}
%
The coefficients in these equations are defined as
\begin{subequations}
\begin{align}
\xi^{(\pm)} &= \gamma(\omega_{-})
\sin^{4}{\theta} \pm \gamma(\omega_{+})\cos^{4}{\theta}\,, \\
\zeta^{(\pm)}_{ab} &= \gamma(\omega_{-})\chi_{ab}(\omega_{-})
\sin^{4}{\theta} \pm \gamma(\omega_{+})\chi_{ab}(\omega_{+})\cos^{4}
{\theta}\,, \\
c^{(0)}_{ab} &= \gamma(\omega_{L})[1 - \chi_{ab}
(\omega_{L})]\sin^{2}{2\theta}\,.
\end{align}
\end{subequations}
Simple analytical expressions for the two-atom steady-state quantum 
dynamics can be obtained if the two-particle interaction is mediated
by the usual vacuum modes of the environmental electromagnetic reservoir,
i.e. $\gamma(\omega_{-})$ = $\gamma(\omega_{+})$ = $\gamma(\omega_{L}) 
\equiv \gamma$. Then, one obtains that
\begin{eqnarray}
x = 2\xi^{(-)}/\xi^{(+)}, \quad y = 0, \quad {\rm and }~~z = [\xi^{(-)}/
\xi^{(+)}]^{2}. \label{small}
\end{eqnarray}
In this case, the diagonal atomic dynamics is independent 
of the inter-atomic separation providing that $\Omega \gg 
\{\gamma\Omega_{ab}, N\gamma\}$. In particular for $\theta = \pi/4$, i.e.,
 exact resonance, we find that $x=y=z=0$, which means that the single-atom
dressed states are equally populated.

\section{Second-order correlation functions and two-particle side-band photon 
correlations}
We now turn to the second-order correlation function of the steady-state 
resonance fluorescence emitted in the three spectral bands. The coherence 
properties of an electromagnetic field, at space-point $\vec R$, can 
be evaluated with the help of the second-order coherence functions:
\begin{align}
g^{(2)}_{mn}(\tau, \vec R_1, \vec R_2) = \frac{\langle a^{+}_{m}(t,\vec R_1) 
\, a^{+}_{n}(t+\tau,\vec R_2) \, a_{n}(t+\tau,\vec R_2) 
\, a_{m}(t,\vec R_1)\rangle}
{\langle a^{+}_{m}(t,\vec R_1) \,a_{m}(t, \vec R_1)\rangle \:
\langle a^{+}_{n}(t, \vec R_2) \,a_{n}(t, \vec R_2)\rangle},
\end{align}
where $a_n^{+} ~(a_n)$ are the photon creation (annihilation) operator 
for modes $n\in \{C,L,R\}$.
The quantity $g^{(2)}_{mn}(\tau)$ can be interpreted as a measure for the 
probability
for detecting one photon emitted in mode $m$ and another photon emitted 
in mode $n$
with delay $\tau$. From now on, all correlation functions
are evaluated for $\tau = 0$, and, for notational simplicity, we 
drop the variable $\tau$.
We further define two Cauchy-Schwarz parameters
\begin{eqnarray}
\label{CSG}
\chi_{L}(\vec R_1, \vec R_2) = \frac{g^{(2)}_{LL}(\vec R_1, \vec R_2)
g^{(2)}_{RR}(\vec R_1, \vec R_2)}{[g^{(2)}_{LR}(\vec R_1, R_2)]^{2}}
\,, \quad 
\chi_{R}(\vec R_1, \vec R_2) = \frac{g^{(2)}_{LL}(\vec R_1, \vec R_2)
g^{(2)}_{RR}(\vec R_1, \vec R_2)}{[g^{(2)}_{RL}(\vec R_1, R_2)]^{2}}\,,
 \end{eqnarray}
which relate the correlation between photons emitted into individual modes
to the cross-correlation between photons emitted into two different
modes. If $\chi_L<1$ or $\chi_R<1$, the respective Cauchy-Schwarz inequalities
are violated~\cite{CSI}.

We now turn to detection in the far-zone limit, and specialize to 
a resonant driving field ($\theta = \pi/4$).
Then the field operators $a_n^{+} ~(a_n)$ entering in $g^{(2)}_{mn}$ 
and corresponding, respectively, to the three spectral modes of the Mollow 
spectrum  can be represented via the 
atomic operators given in Eq.~(\ref{s_op}). With the help of 
Eqs.~(\ref{H0}-\ref{small}), and in Born-Markov and secular approximation,
one arrives at the following expressions for the two-particle 
photon-correlations:
\begin{subequations}
\label{noVi}
\begin{eqnarray}
g^{(2)}_{CC}(\vec R_{1},\vec R_{2}) &=& 1 + \cos{\delta_{1}}\cos{\delta_{2}} 
\,,  \\
g^{(2)}_{LL}(\vec R_{1},\vec R_{2}) &=& 
g^{(2)}_{RR}(\vec R_{1},\vec R_{2}) = \frac{1}{2}[1 + \cos(\delta_{1}-\delta_{2})]
\,,  \\
g^{(2)}_{LR}(\vec R_{1},\vec R_{2}) &=&
g^{(2)}_{RL}(\vec R_{1},\vec R_{2}) =  \frac{1}{2}[3 + \cos(\delta_{1}+\delta_{2})] 
\,, \\
g^{(2)}_{CX}(\vec R_{1},\vec R_{2}) &=& 
g^{(2)}_{XC}(\vec R_{1},\vec R_{2}) = 1 \qquad \textrm{ for } X\in \{L,R\}\,.
\end{eqnarray}
\end{subequations}
%
Here we used that in the resonant strong-field limit one has $\langle R_{zj}\rangle_{s}=
\langle R^{(j)}_{12}\rangle_{s}=\langle R^{(j)}_{21}\rangle_{s}=0$ and
$\langle R_{zi}R_{zj}\rangle_{s}=\langle R^{(i)}_{21}R^{(j)}_{12}\rangle_{s}=0$,
where $i \not=j \in \{a,b\}$ denote one of the two atoms.
The parameters $\delta_d$ ($d\in\{1,2\}$) are related to the detection angles
$\alpha_d$ by $\delta_d = k\, r_{ab}\,\cos\alpha_d$, where we have assumed
a planar geometry of atomic sample and observation detections.
%
%
\begin{figure}[t]
\includegraphics[width=7.8cm]{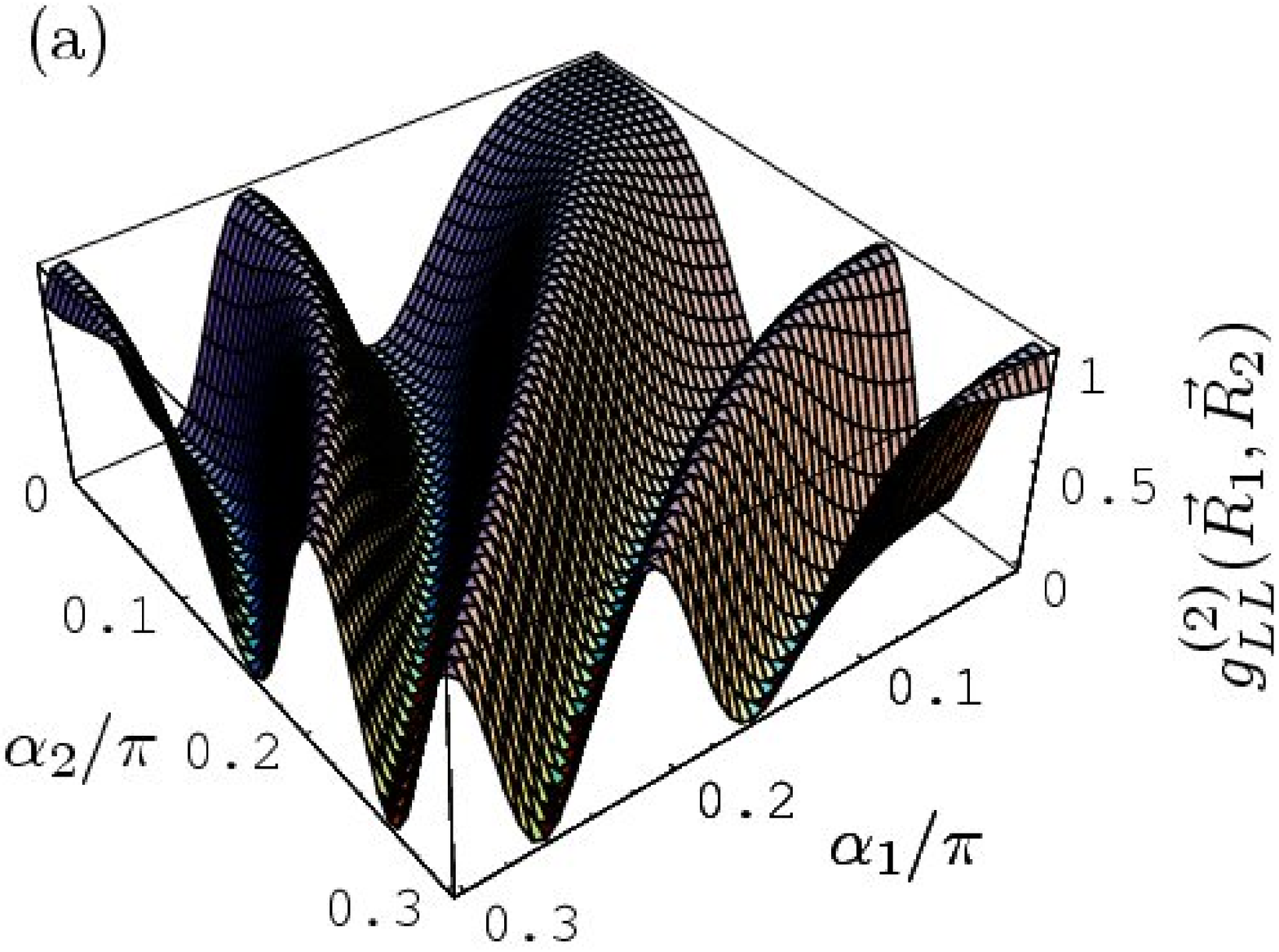}
\hspace*{0.5cm}
\includegraphics[width=7.8cm]{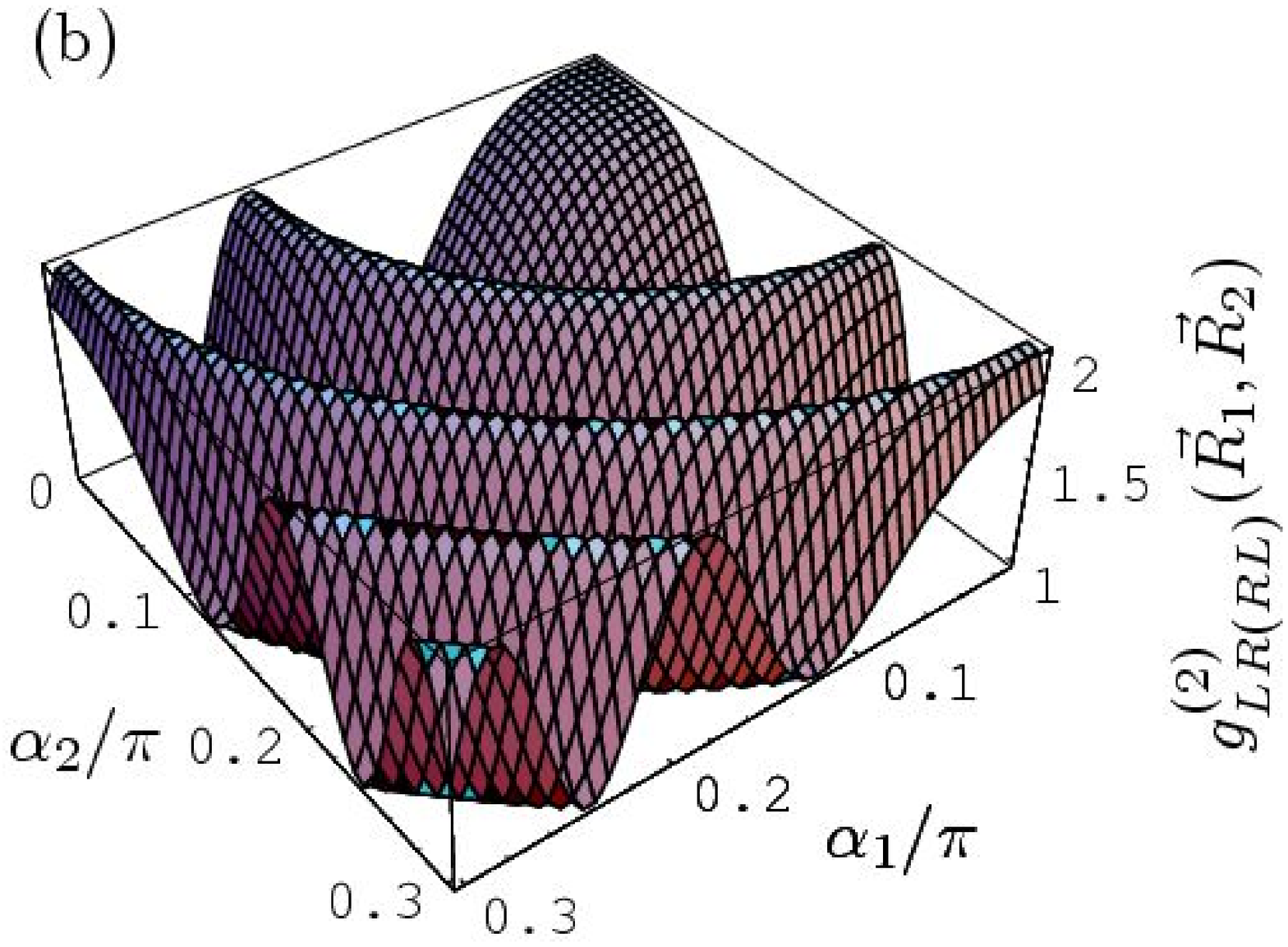}
\caption{\label{fig-2}
(a) Second-order side-band photon correlations 
$g_{LL(RR)}^{(2)}(\vec{R}_1,\vec{R}_2 )$ 
and (b) cross-correlations 
$g_{LR(RL)}^{(2)}(\vec{R}_1,\vec{R}_2 )$
as function of 
the detector positions $\alpha_{1},\alpha_{2}$.
The interatomic distance is $r_{ab}/\lambda=5$, and the 
number of atoms is $N=2$.
}
\end{figure}

Already the structure of Eqs.~(\ref{noVi}) reveals interesting insight in the
nature of the light emitted in the various spectral bands.
From the properties of the $\cos$ function it follows that
$0\leq g^{(2)}_{CC}(\vec R_{1},\vec R_{2}) \leq 2$, and 
$0\leq g^{(2)}_{LL}(\vec R_{1},\vec R_{2})=g^{(2)}_{RR}(\vec R_{1},\vec R_{2}) \leq 1$,
and
$1 \leq g^{(2)}_{LR}(\vec R_{1},\vec R_{2}) =
g^{(2)}_{RL}(\vec R_{1},\vec R_{2}) \leq 2$.
The cross-correlations involving the central spectral band are always unity
and do not exhibit a dependence on the detection position.
This immediately fixes the possible photon statistics of the emitted light. 
We observe a tendency to have pairs of photons emitted in the central
band or pairs where one of the photons is in either sideband (left and
right).
In general, for these correlation functions, values below unity denote
sub-poissonian
light statistics, a value of units indicates poissonian statistics, and
values above unity stand for super-poissonian light statistics.

An example for the second-order sideband photon correlation functions
$g_{LL}^{(2)}$ and $g_{RR}^{(2)}$ as a function of the detection angles
$\delta_1, \delta_2$ is shown in Fig.~\ref{fig-2}(a). 
The spatial interference fringes of the second-order correlation function are
clearly visible, and the range of possible values 
for $g_{LL}^{(2)}$ and $g_{RR}^{(2)}$ shows that both poissonian and
sub-poissonian photon statistics can be generated at the sideband 
frequencies.
Fig.~\ref{fig-2}(b) shows the corresponding results for the
second-order cross-correlations $g_{LR}^{(2)}$ and $g_{RL}^{(2)}$.
Here, the oscillatory structure is different, and values from 
1 to 2 are obtained for the cross-correlations. Therefore
poissonian as well as super-poissonian light statistics
can be generated in the cross-correlations.
In general, sub-Poissonian and Poissonian
light statistics can be generated in all three spectral lines,
and super-Poissonian light statistics is possible in the central spectral band
second-order correlation function $g_{CC}^{(2)}$ and in the
cross-correlations, see, e.g., Eq.~(\ref{noVi}).

We now discuss the second-order spatial interference
of the light emitted in the central frequency band around $\omega_L$
for a specific detection scheme.
We assume identical detection angles $\delta_{1}=\delta_{2}\equiv \delta$
and detection positions $\vec R_1 = \vec R_2 \equiv \vec R$,
which corresponds, e.g., to a medium sensitive to two-photon 
exposure~\cite{double-res}. In the following, we discuss the
interference fringe resolution of this setup, which is of relevance
to applications in lithography, where the general aim is to create 
structures as small as possible.
In the strong-field limit $\Omega/\gamma \gg 1$, 
from Eq.~(\ref{noVi}) we find $g^{(2)}_{CC}(\vec R) = 1+\cos^{2}{\delta}$. 
In the weak field case $\Omega/\gamma <1$ without spectral band 
separation, however, one finds $g^{(2)}(\vec R) = [s/(s + \cos{\delta})]^{2}$
with $s=1 + 2(\Omega/\gamma)^{2}$~\cite{int2}. 
Comparing these two results, one finds that simply increasing the driving
field strength effectively doubles the spatial fringe resolution in this
setup around the central frequency $\omega_L$.
%

%
\begin{figure}[t]
\includegraphics[height=6.4cm]{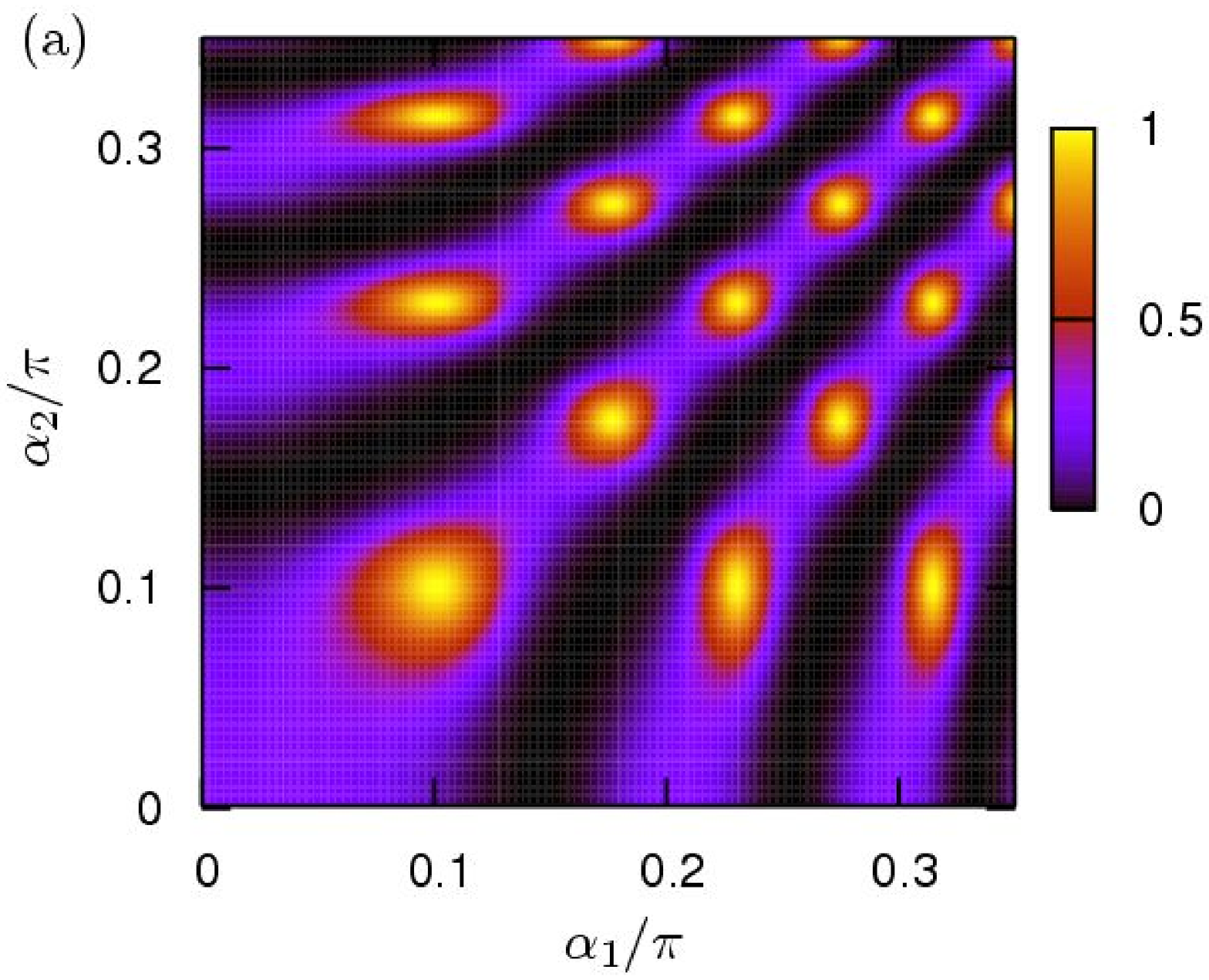}
\hspace*{0.5cm}
\includegraphics[height=6.4cm]{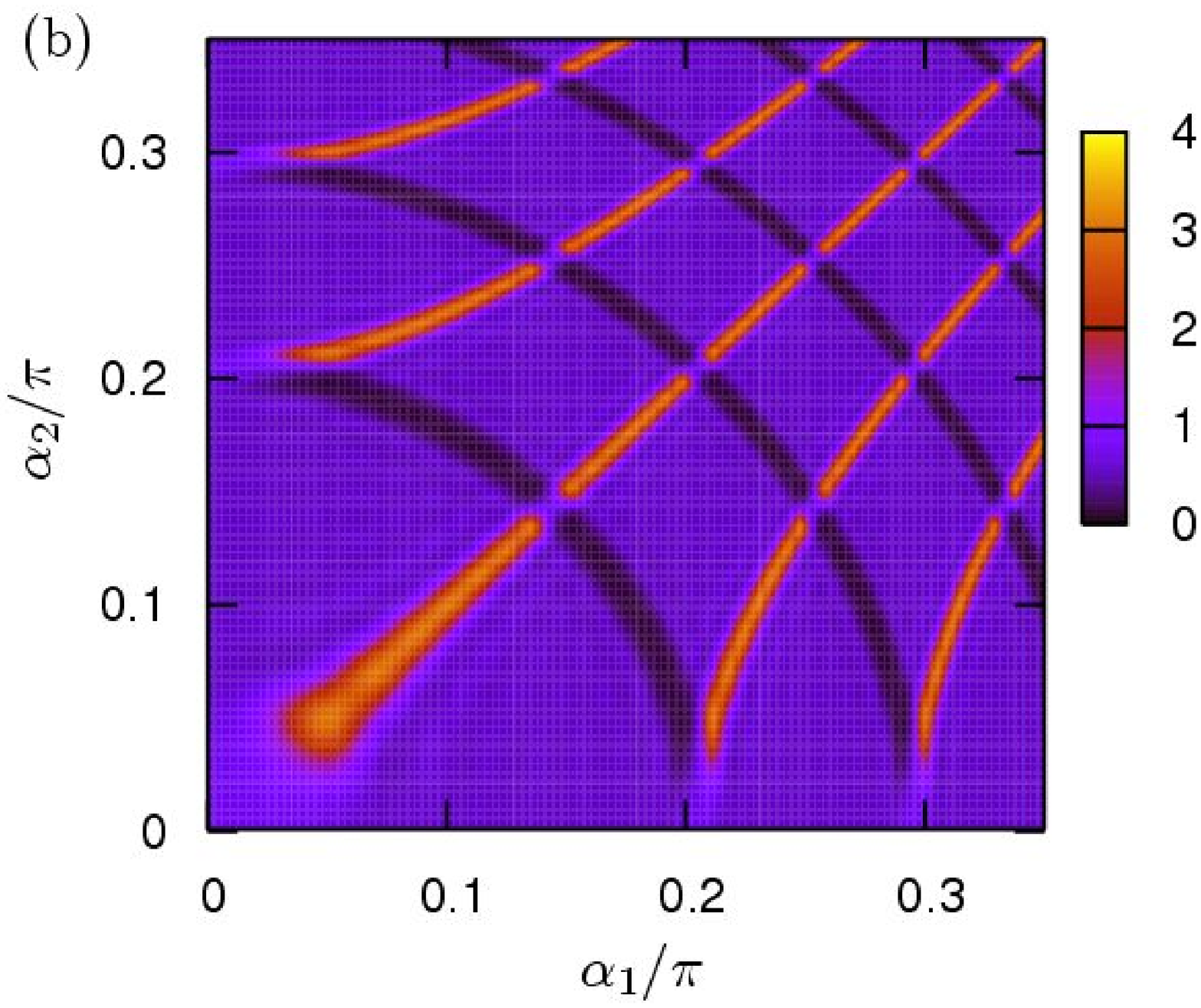}
\caption{\label{fig-4} The Cauchy-Schwarz parameter $\chi_{L}=\chi_{R}$
as function of detector positions $\alpha_{1},\alpha_{2}$. 
Here, $r_{ab}/\lambda=5$. The number of atoms is (a) $N=2$ and (b) $N=8$.}
\end{figure}

After the discussion of the central spectral band, we now turn to the
sideband frequencies around $\omega_L\pm 2\Omega$.
Again, we consider detection in the far zone limit, but at this time
the two detectors are at different positions.
Under this conditions, the Cauchy-Schwarz parameters in Eq.~(\ref{CSG})
evaluate to
\begin{eqnarray}
\chi_{L}=\chi_{R}= \biggl[\frac{1+\cos(\delta_{1}-\delta_{2})}
{3+\cos(\delta_{1}+\delta_{2})}\biggr]^{2}. \label{CS}
\end{eqnarray}
%
This results is shown in Fig.~\ref{fig-4}(a) for an interparticle distance
$r_{ab}=5\lambda$ and two atoms ($N=2$). One may easily observe that 
for a large range of detector positions $\delta_1, \delta_2$, the
Cauchy-Schwarz inequalities are violated, i.e., $\chi_{L} = \chi_{R} < 1$.
These results for $\chi_{L}$ and $\chi_{R}$ can be easily understood by 
inspecting Eq.~(\ref{CSG}) and Fig.~\ref{fig-2}. The Cauchy-Schwarz parameters
are given by the ratio of the product of the sideband second-order 
photon correlations $g^{(2)}_{LL}(\vec R)$ and $g^{(2)}_{RR}(\vec R)$
to the cross correlation $g^{(2)}_{LR}(\vec R)$ or $g^{(2)}_{RL}(\vec R)$
squared. Therefore, both oscillatory structures in Fig.~\ref{fig-2} are
combined to give the result in Fig.~\ref{fig-4}(a).
It is interesting to note that it is not possible to distinguish
whether the first photon is emitted on the left or on the right
sideband, since the two different cross-correlations
$g^{(2)}_{LR}(\vec R)$ and $g^{(2)}_{RL}(\vec R)$ are equal.

\begin{figure}[t]
\includegraphics[height=6.4cm]{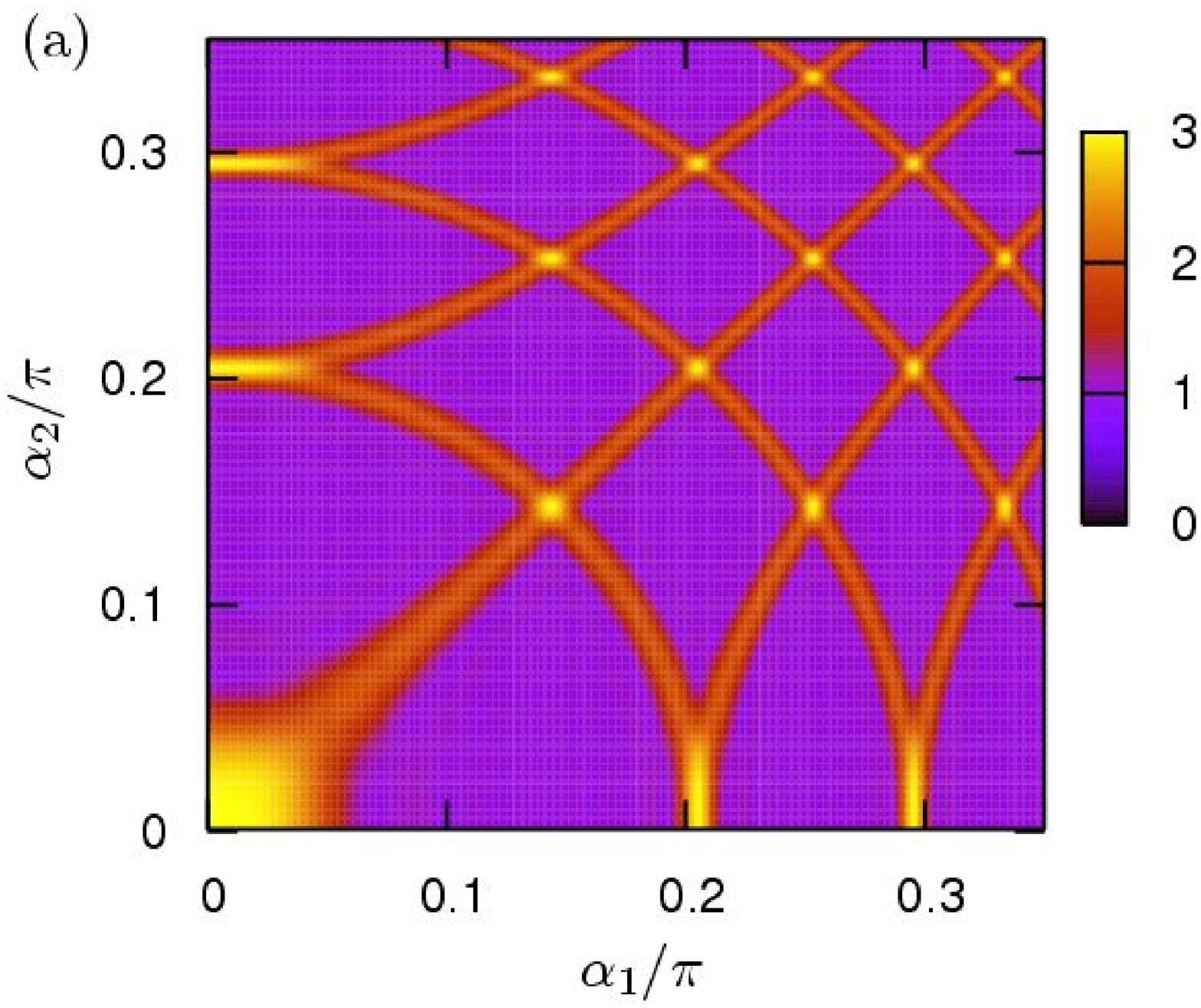}
\hspace*{0.5cm}
\includegraphics[height=6.4cm]{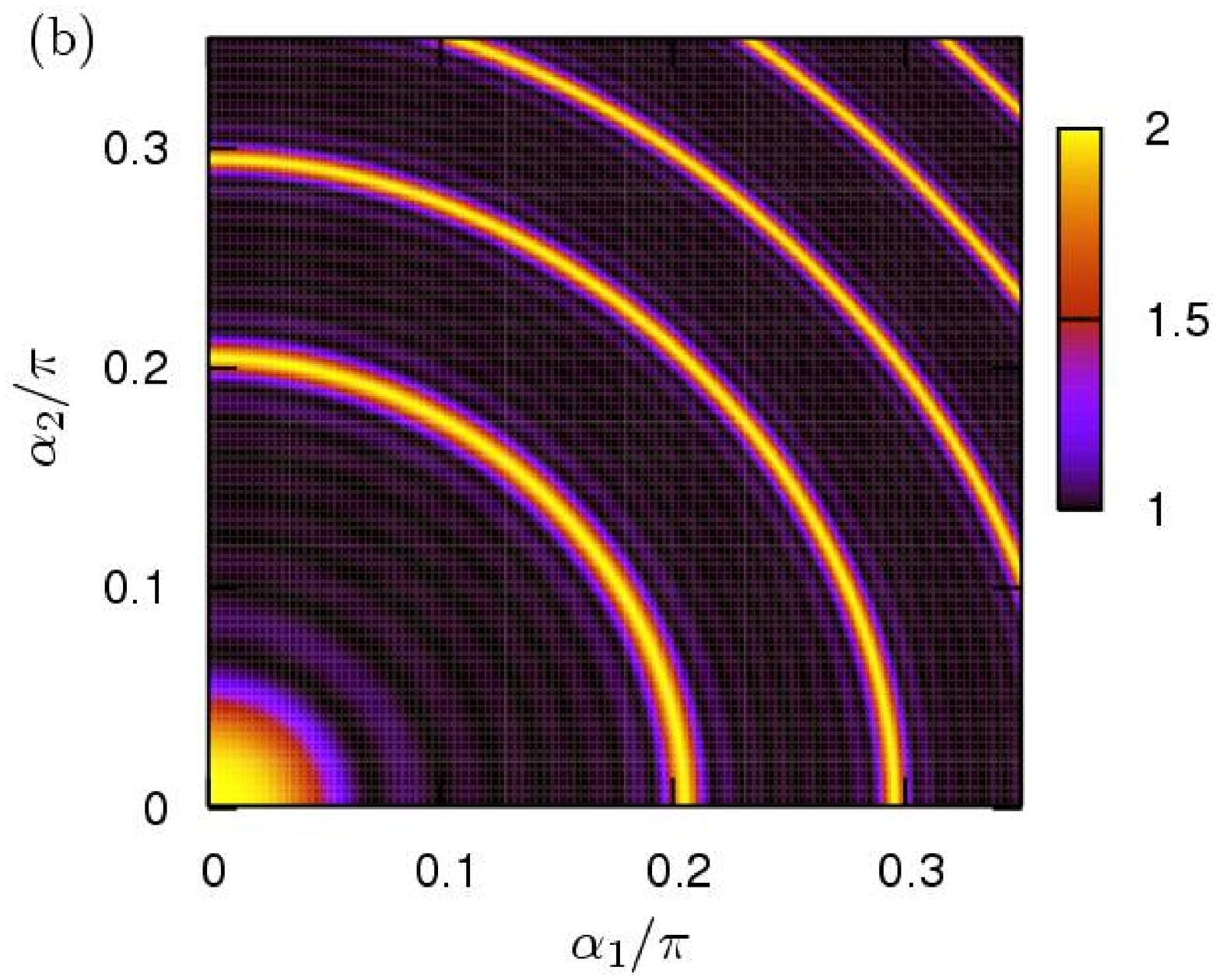}
\caption{\label{fig-6}
(a) Second-order side-band photon correlations 
$g_{CC}^{(2)}(\vec{R}_1,\vec{R}_2 )$ 
and (b) $g_{LR}^{(2)}(\vec{R}_1,\vec{R}_2 )
= g_{RL}^{(2)}(\vec{R}_1,\vec{R}_2 )$ 
as function of 
the detector positions $\alpha_{1},\alpha_{2}$.
The interatomic distance is $r_{ab}/\lambda=5$, and the 
number of atoms is $N=8$.
}
\end{figure}

\section{\label{multi}Multi-particle structures}
In this final part, we discuss the case of $N$ independent two-level 
emitters that are uniformly distributed in a regular chain.
Thus the inter-particle separations between the neighbor particles
is assumed uniform and equal to $r_{0}$. As before, the atoms
are driven by a strong laser field, $\Omega/\gamma\gg 1$.
We further assume detection in the far-zone limit, that is, 
the linear dimension of the chain $L=(N-1)r_{0}$ is much smaller 
than the distances between chain and detectors $|\vec R_{1}|$ and $|\vec R_{2}|$.
Then the second-order correlation functions 
for the light emitted in the central- and the sidebands as well as to the 
cross-correlations can be evaluated to give
\begin{subequations}
\begin{eqnarray}
g^{(2)}_{CC}(\vec R_{1},\vec R_{2}) &=& 1 - \frac{2}{N}+\frac{1}{N^{2}}[\phi(\delta_{1}+\delta_{2})
+ \phi(\delta_{1}-\delta_{2})] 
\,,  \\
g^{(2)}_{LL}(\vec R_{1},\vec R_{2}) &=& g^{(2)}_{RR}(\vec R_{1},\vec R_{2})
= 1 - \frac{2}{N} + \frac{1}{N^{2}}\phi(\delta_{1}-\delta_{2})
\,,  \\
g^{(2)}_{LR}(\vec R_{1},\vec R_{2}) &=& 
g^{(2)}_{RL}(\vec R_{1},\vec R_{2})
= 1 + \frac{1}{N^{2}}\phi(\delta_{1}+\delta_{2}) \,,  \\
g^{(2)}_{CX}(\vec R_{1},\vec R_{2}) &=& 
g^{(2)}_{XC}(\vec R_{1},\vec R_{2}) = 1 \qquad \textrm{ for } X\in \{L,R\}\,,
\label{cf}
\end{eqnarray}
\end{subequations}
where $\phi(\delta)=\sin^{2}(N\delta/2)/\sin^{2}(\delta/2)$.

In the multi-particle case, the Cauchy-Schwarz parameters  Eq.~(\ref{CSG})
for the side-band photon correlations are given by
the following expression
\begin{eqnarray}
\label{CSN2}
\chi_{L}=\chi_{R}= \biggl[\frac{N^{2} -2N + \phi(\delta_{1}-\delta_{2})}{N^{2} +  
\phi(\delta_{1}+\delta_{2})}\biggr]^{2}. \label{CSN}
\end{eqnarray}
%
Figure~\ref{fig-4}(b) shows this Cauchy-Schwarz parameter for a
linear chain of $N=8$ atoms. In this figure, the Cauchy-Schwarz
parameters assume values from below unity up to 4.
In the following, we discuss two special cases to demonstrate this 
more clearly. First, consider  a two-photon detector with
$\delta_{1}=\delta_{2} \equiv \delta$.
Then Eq.~(\ref{CSN2}) reduces to 
\begin{align}
\chi_{L}=\chi_{R}= \left ( \frac{2N(N-1)}{N^{2} + \phi(2\delta)} \right )^{2} 
\end{align}
which for $N\gg 1$ and $\delta = n\pi~(n=0,1,\cdots)$
tends to $\chi_{L(R)} \to 1$ while for $\delta = \pi(1+2n)/2$ we have 
$\chi_{L(R)} \to 4$. If, for instance, $\delta_{1}(\delta_{2})=0$  and 
$\delta_{2}(\delta_{1})\equiv \delta$ with $N \gg 1$ then $\chi_{L(R)}$ goes 
to unity as $\chi_{L(R)}=[1-1/N]^{2}$ when $\delta = n\pi~(n=0,1,\cdots)$ 
and to $\chi_{L(R)}=[1-2/N]^{2}$ for $\delta = \pi(1+2n)/2$. Thus, violation
of CSI is likely to occur for moderate atomic structures. 

Fig.~\ref{fig-6}(a) shows the central band second order correlation function
$g^{(2)}_{CC}(\vec R_{1},\vec R_{2})$ for $N=8$ atoms versus the detection
angles $\alpha_1, \alpha_2$. Comparing this figure with Fig.~\ref{fig-4}(b),
it is apparent that the dark regions in 
Fig.~\ref{fig-4}(b) that correspond to a super-poissonian photon
emission statistics for a pair of photons from the left and the 
right spectral sideband [see Fig.~\ref{fig-6}(b)] are regions of high correlation in 
Fig.~\ref{fig-6}(a) as well. Thus, in these detection directions,
super-poissonian statistics is observed in the central spectral band
and the cross-correlations of the sideband photons.

Finally we note that the properties of the correlation functions 
found here are quite different from those obtained in a Dicke-type
sample~\cite{CS_B,dicke2,CS_mek}. For example, Dicke-type samples
do not exhibit the spatial dependence originating from the regular
structure discussed here, since the Dicke model involves the assumption
that all atoms interact with the electromagnetic fields with the same
phase. In particular, the first-order correlation function
in our present model is linear in the number of atoms $N$, 
whereas it is proportional to the number of atoms squared
in the Dicke model. The unnormalized second-order correlation function
is proportional to $N^4$ and does not depend on the detection direction
in Dicke-type samples. For regular structures, the unnormalized
second-order correlation function depends at most on $N^2$ for particular
detection positions. This detector position dependence arises from
the geometrical phase factors $\exp(i\vec k \cdot \vec r_{ab})$, and not from collectivity.
Constructive interfere of these phase factors gives rise to the dependence on $N^2$
for particular observation directions.

\section{Summary and Conclusion}
In summary, the intensity-intensity correlations of
photons scattered by a strongly pumped linear atomic
structure was investigated in detail. 
For this, the spectrum of the scattered light was separated
into different spectral bands which we treated independently.
In the central band at the driving laser field frequency,
the resonant two-particle second-order interference fringes 
for two-photon detection have double resolution in the 
strong-field case as compared to the weak-field pattern. 
We found violation of the spatial Cauchy-Schwarz inequalities for 
photons emitted into the spectral sidebands, and that it 
is impossible to predict which photon is emitted first. 
Multiparticle atomic structures significantly enhance 
the intensity of the emitted correlated photons, and exhibit 
properties different from those in Dicke-type multiparticle
samples. 



\end{document}